\documentclass[pre,aps,showpacs]{revtex4}

\usepackage{amssymb}
\usepackage{amsmath}
\usepackage{latexsym}

\usepackage[dvips]{graphicx}

\newcommand{\sikib}{\begin{eqnarray}}
\newcommand{\sikie}{\end{eqnarray}}
\newcommand{\sikibnon}{\begin{eqnarray*}}
\newcommand{\sikienon}{\end{eqnarray*}}

\newcommand{\ptot}{\Delta p_{\perp}}
\newcommand{\pup}{\Delta p_{\uparrow}}
\newcommand{\pdown}{\Delta p_{\downarrow}}
\newcommand{\iup}{\Delta I_{\uparrow}}
\newcommand{\idown}{\Delta I_{\downarrow}}
\newcommand{\etot}{\left< e \right>}
\newcommand{\tup}{\left< t_{\uparrow} \right>}
\newcommand{\tdown}{\left< t_{\downarrow} \right>}

\newcommand{\pd}{\partial}

\begin{document}

\title{Steady State Thermodynamic Functions for Radiation Field}

\author{Hiromi Saida}
\email{saida@daido-it.ac.jp}

\affiliation{Department of Physics, Daido Institute of Technology\\
             Takiharu-cho 10-3, Minami-ku, Nagoya 457-8530, Japan}

\baselineskip=5mm

\begin{abstract}

Several attempts to construct the thermodynamical formulation for non-equilibrium steady states have been made along the context by Oono and Paniconi, called the steady state thermodynamics (SST). In this paper we study the SST for radiation field. Consider a cavity in a large black body, and put an another black body in it. Let the temperature of the outer black body be different from that of the inner one, and keep their temperatures unchanged. Then the radiation field between the black bodies are retained in a non-equilibrium steady state with a stationary energy flow. (The inner black body may be considered as a representative of a black hole with the Hawking radiation, and the cavity be the universe including the black hole.) According to the total momentum carried by photons par a unit time in passing through a virtual unit area, the pressure for the SST is defined. Then, taking this pressure as the basis, we propose the definitions of steady state thermodynamic functions and state variables satisfying the plausible properties: the convexity/concavity of thermodynamic functions, the intensivity/extensivity of state variables, the Gibbs-Duhem relation and the Legendre transformation among thermodynamic functions. 

\end{abstract}


\maketitle

\section{Introduction}\label{sec-intro}

Several attempts to construct the thermodynamical formulation for non-equilibrium steady states have been made along the context by Oono and Paniconi \cite{ref-op}, called the steady state thermodynamics (SST) \cite{ref-sst}. In this paper we study the SST for radiation field. The reason we consider the radiation field is the future expected application to black hole physics. The black hole is the spacetime region of extremely strong gravity so that no matter can escape from the region in the context of classical physics (the theory of general relativity) \cite{ref-he}. However it is well known that the quantum effect of the matter field on the black hole spacetime makes the black hole emit a thermal radiation from its event horizon (Hawking radiation) \cite{ref-bh}. This means that the black hole can evaporate due to the energy loss by the Hawking radiation. While it is valid to assume that the black hole itself is in an equilibrium state, but the radiation field emitted outside the black hole should be considered in a non-equilibrium state with the energy flow by the Hawking radiation. However, so far, the matter fields surrounding the black hole have been assumed to be in an equilibrium state, and it has been impossible to trace some details of the evaporation process with changing the black hole's temperature and entropy at once \cite{ref-bh}. Therefore we can not avoid considering a non-equilibrium state of the radiation field for a total understanding of the black hole evaporation. 

Although our final aim is the application to black hole physics, we set the black hole physics aside and concentrate on the system which can be designed for a laboratory experiment. Consider a cavity in a large black body, and put an another black body in the cavity. (The inner one is a representative of the black hole with the Hawking radiation.) Let the temperature of the outer black body $T_{out}$ be different from that of the inner one $T_{in} (\neq T_{out})$, and keep their temperatures unchanged. Then the radiation field between the black bodies are retained in a non-equilibrium steady state with a stationary energy flow. Hereafter we abbreviate the non-equilibrium steady state with a stationary energy flow to the steady state. We treat this radiation field throughout this paper and search the consistent definitions of steady state thermodynamic functions and state variables for the radiation field within the phenomenological framework. 

\begin{figure}[t]
 \begin{center}
  \includegraphics[height=30mm]{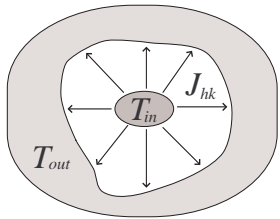}
 \end{center}
\caption{Non-equilibrium steady state with a stationary energy flow for a radiation field. A black body of temperature $T_{in}$ is put in the cavity made in a large black body of temperature $T_{out}$. The house-keeping heat rate $J_{hk}$ retains the radiation field in a steady state.}
\label{pic.sys.1}
\end{figure}

Further we do not consider any statistical and quantum fluctuations in this paper, and assume that both of black bodies emit always thermal radiations of their temperatures respectively without respect to the temperature difference. Then the energy $J_{hk}$ which we should supply to the radiation field par a unit time in order to retain the steady state is given by
\sikib
 J_{hk} = \sigma \left( T_{in}^4 - T_{out}^4 \right) A_{in} \, ,
 \label{eq-hk}
\sikie
where $\sigma$ is the Stefan-Boltzmann constant and $A_{in}$ is the surface area of the inner black body. This $J_{hk}$ is the {\it house-keeping heat rate} explained below.

Along the context of the original consideration of the SST \cite{ref-op}, we need to separate the total heat supplied to the radiation field during a change of steady state into two parts, {\it house-keeping heat} and {\it excess heat}. The house-keeping heat is the necessary heat supplied to the radiation field in order to make it stay at one steady state (without any change of the state). The longer the time interval for keeping the steady state becomes, the larger the house-keeping heat we should supply. So it is useful to consider the house-keeping heat {\it rate} supplied par a unit time, which is given by eq. (\ref{eq-hk}). In changing the state of the radiation field from one steady state to another one, the total heat {\it rate} $J_{tot}$ supplied to the radiation field at each moment during this change depends not only on the path of the change of steady state in the state space (of steady states), but also on the speed and acceleration of the change along the path in the state space, the geometrical shape of the system and the temperatures of black bodies. Therefore it is necessary to supply more (less) energy than $J_{hk}$ to raise (lower) the temperature difference $T_{in} - T_{out}$, and consequently $J_{tot}$ is generally different from $J_{hk}$ at each moment during the change of steady state. Then the excess heat {\it rate} $J_{ex}$ is defined by $J_{ex} = J_{tot} - J_{hk}$, which is the part of $J_{tot}$ required to drive the change of steady state of the radiation field. Note that we find $J_{ex} = J_{tot}$ for equilibrium cases ($J_{hk} = 0$). The central spirit of the SST is the hypothesis that the excess heat in the SST plays the role of the total heat in equilibrium thermodynamics to define the SST entropy which specifies the reversible/irreversible processes for the steady states. Consequently we postulate that the phenomenological formulation of the steady states (SST) for the radiation field exists after extracting the effects of $J_{hk}$ from various quantities, and that the SST has the same mathematical form as the four laws of the equilibrium thermodynamics using the excess heat in place of the total heat of equilibrium cases. Note that this postulation denotes the following three requirements. The first requirement is the existence of the state variables which characterise each steady state completely and uniquely. The second one is that the SST formulation includes the state variables describing the non-equilibrium order which vanish in equilibrium cases. Finally, the third requirement is to keep the basic properties of state variables unchanged: the convexity/concavity of thermodynamic functions, the intensivity/extensivity of state variables, the Gibbs-Duhem relation and the Legendre transformation among thermodynamic functions.

In section \ref{sec-press}, the definition of SST pressure are proposed, whose value is operationally determinable by laboratory experiments. Referring to this SST pressure in section \ref{sec-free}, we suggest the consistent definitions of SST free energy and the other state variables after introducing the intensive state variable specifying the non-equilibrium order. Section \ref{sec-others} is devoted to show the consistency of our definitions, but the SST second law is left to be confirmed in the future work. Summary and discussions are located in section \ref{sec-sd}.

\section{SST pressure}
\label{sec-press}

For constructing the SST for the radiation field, we should make state variables be determinable operationally. Therefore we search the definition of the pressure for the SST (SST pressure) using the total momentum carried by photons passing through a unit area par a unit time. Then an operational method for measuring the SST pressure is given according to the definition. But before considering the steady state, we start with reviewing the equilibrium case.

\subsection{For equilibrium case}

Let the temperatures of black bodies be equal $T_{in} = T_{out} \equiv T_{eq}$, then the radiation field between them is in an equilibrium state of the pressure $P_{eq}(T_{eq}) = (4 \sigma/3 c) \, T_{eq}^{\, 4}$, where $c$ is the speed of light. In the region filled with the radiation field (between the black bodies), image a virtual closed smooth surface $C$ surrounding the inner black body. Then, as is well known, the relation $P_{eq} = \ptot$ holds, where $\ptot$ is the total normal component of the momentum to $C$ carried by photons passing through $C$ par a unit area and a unit time. We divide it as $\ptot = \pdown + \pup$, where $\pdown$ is the part due to the photons crossing $C$ from the exterior region of $C$, and $\pup$ is due to the photons crossing $C$ from the interior region of $C$. Note that, while all photons contributing to $\pdown$ are emitted at the surface of the cavity, but the photons contributing to $\pup$ consist of ones emitted at the inner black body's surface and the others emitted at the cavity's surface. So further we can divide as $\pup = \pup^{in} + \pup^{out}$, where $\pup^{in}$ is the component emitted at the inner black body's surface and $\pup^{out}$ is that emitted at the cavity's surface. Therefore we find
\sikib
 P_{eq} = \pdown + \left( \, \pup^{in} + \pup^{out} \, \right) \, .
\label{eq-Peq.1}
\sikie

Next let us place a very thin wall on $C$ whose inner and outer surfaces are covered with perfect mirrors, and keep the temperature of the mirror wall be absolutely zero so that the mirror wall emits no photon and just reflects the photons emitted at the other black bodies. Then the radiation field in the exterior region of the mirror wall is in an equilibrium state with the pressure $P_{eq}$, and the radiation field in the interior region is also in the same state. For this case, as is well known, we find $P_{eq} = \idown = \iup$, where $\idown$ is the normal component (to $C$) of the impact given to a unit area on the mirror wall par a unit time by the photons in the exterior region, and $\iup$ is the impact by the photons in the interior region. Further, due to the perfectly elastic collision of photons at the mirror wall, the relations $\idown = 2 \pdown$ and $\iup = 2 \pup$ hold. Consequently we obtain
\sikib
 \pdown = \pup^{in} + \pup^{out} = \frac{1}{2} \, P_{eq}
\label{eq-Peq.2}
\sikie

\begin{figure}[t]
 \begin{center}
  \includegraphics[height=35mm]{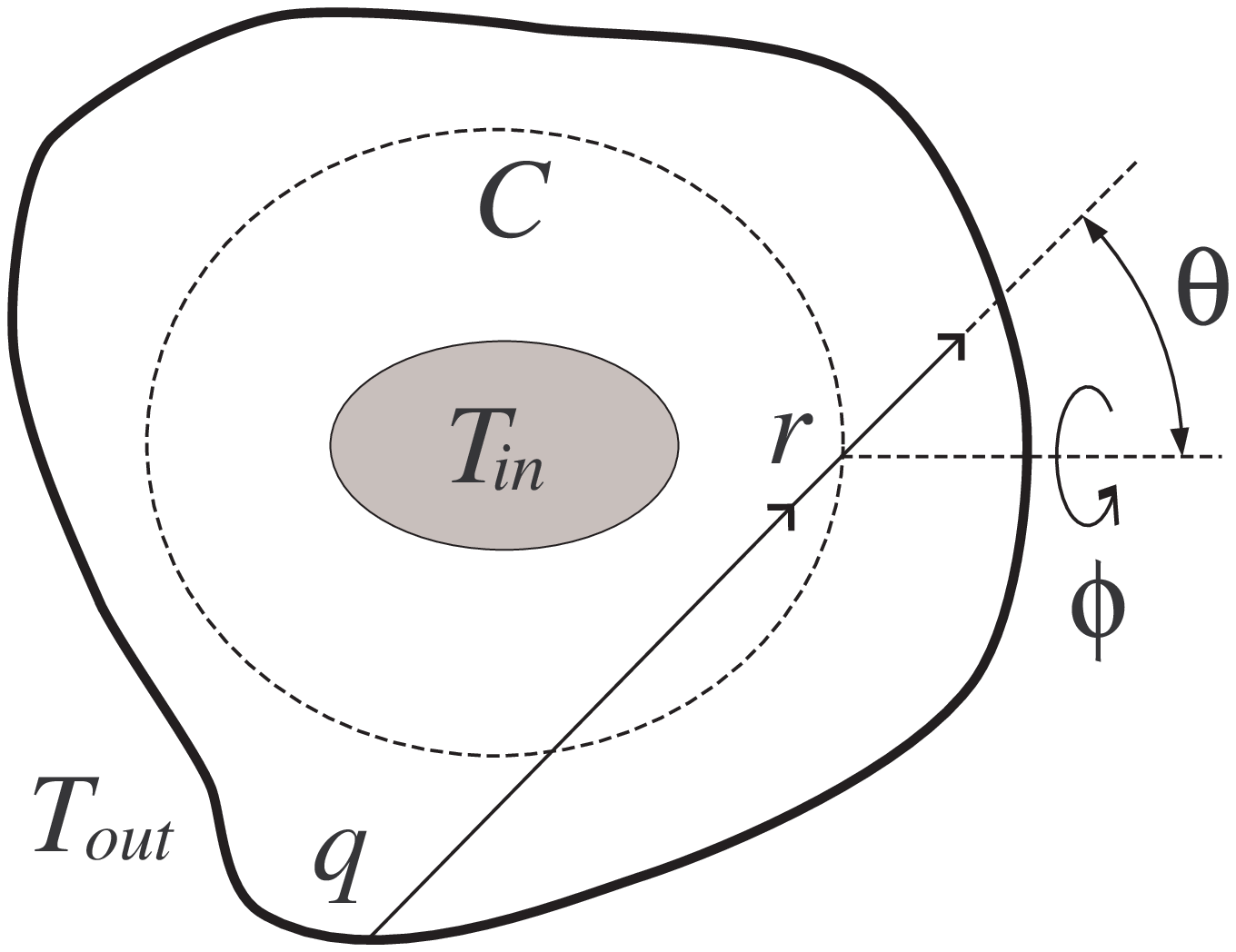}
 \end{center}
\caption{The photon which contributes to $\pup^{out}$.}
\label{pic.mom.1}
\end{figure}

Here, after removing the mirror wall, let us calculate $\pup^{out}$ explicitly using the black body spectrum of photons emitted at the cavity's surface. Consider a photon emitted at a point $q$ on the cavity's surface to a point $r$ on the closed surface $C$ as shown in figure \ref{pic.mom.1}. The magnitude of the momentum of this photon is $\hbar \omega/c$, where $\omega$ is the frequency of the photon. The normal component of this momentum to the surface $C$ at $r$ is $(\hbar \omega/c)\, \cos\theta$, where $\theta$ is the angle between the momentum and the normal direction to $C$. Note that, because of the inner black body's shielding, no photon comes to $r$ from the cavity's surface for some range of the angles $\theta$ and $\phi$, where $\phi$ is the zenithal angle with respect to the normal direction to $C$ at $r$. We introduce a function $\epsilon(\theta , \phi)$ where $\epsilon = 0$ for the range of the angles $\theta$ and $\phi$ with no coming photon to $r$ and $\epsilon = 1$ for the other range of the angles. Using the black body spectrum, the number of photons coming to $r$ from $q$ par a unit area on $C$, a unit time, a unit frequency and a unit solid angle is given by
\sikibnon
 \epsilon(\theta ,  \phi) \,
  \frac{1}{4 \pi^3 c^2} \, \frac{\omega^2}{\exp(\beta \hbar \omega) -1} \,
   \cos\theta \, ,
\sikienon
where $\beta = 1/k_B T_{eq}$ and $k_B$ is the Boltzmann constant. Consequently we obtain 
\sikib
 \pup^{out} &=&
   \int_0^{\pi/2} \sin\theta d\theta \int_0^{2 \pi} d\phi \, \int_0^{\infty} d\omega \,
   \epsilon \, \frac{1}{4 \pi^3 c^3} \,
   \frac{\hbar \omega^3}{\exp(\beta \hbar \omega) - 1} \,
   \cos^2\theta     \nonumber \\
 &=&
  \left[ \,
   \frac{k_B^{\,4}}{4 \pi^3 c^3 \hbar^3} \, 
    \int_0^{\pi/2} d\theta \int_0^{2 \pi} d\phi \,
    \epsilon \, \sin\theta \, \cos^2\theta \,
   \int_0^{\infty} dx \, \frac{x^3}{e^x - 1} \,
  \right] \, T_{eq}^{\, 4}     \nonumber \\
 &\equiv&
  \alpha \, T_{eq}^{\, 4} \, ,
\label{eq-Peq.3}
\sikie
where $\sin\theta$ arises with the integral about the solid angle around $r$, $x = \beta \hbar \omega$ and $\alpha$ is the factor determined by the followings: the geometrical shape of the cavity and the inner black body, the configuration of the inner black body in the cavity, the geometrical shape of $C$ and the choice of the point $r$ on $C$
\footnote{Since the photons contributing to $\pup^{out}$ should pass through $r$ from the interior region of $C$, the integral range of $\theta$ is $0 \to \pi/2$.}. 
From eq. (\ref{eq-Peq.2}), we find $\pup^{in} = P_{eq}/2 - \pup^{out}$. This means that $\pup^{in}$ has the complicated dependence on the geometry of the system as $\pup^{out}$ does. But note that eq. (\ref{eq-Peq.2}) also denotes the summation $\pup^{in} + \pup^{out}$ has no such a complicated dependence and is determined only by the temperature $T_{eq}$
\footnote{Indeed, $\pup^{in}$ is given by eq. (\ref{eq-Peq.3}) with replacing $\epsilon$ by $1 - \epsilon$. Further $\pdown$ is given by eq. (\ref{eq-Peq.3}) with setting $\epsilon = 1$ for all range of the angles, $0 < \theta < \pi/2$ and $0 < \phi < 2 \pi$.}.

\subsection{For steady state}

Let us proceed to searching the definition of the SST pressure. Because photons behave as collisionless particles, the relation (\ref{eq-Peq.1}) holds for the steady state with replacing the temperatures appropriately for each term (and with recalling the closed surface $C$), 
$\ptot(T_{in} , T_{out}) = \pdown(T_{out}) + ( \, \pup^{in}(T_{in}) + \pup^{out}(T_{out}) \, )$, 
where the dependence on the temperature is explicitly presented. Then via eq. (\ref{eq-Peq.2}) and (\ref{eq-Peq.3}), we obtain
\sikib
 \ptot(T_{in} , T_{out})
 &=&
   \frac{1}{2} \, \left( \, P_{eq}(T_{in}) + P_{eq}(T_{out}) \, \right)
  - \left( \, \pup^{out}(T_{in}) - \pup^{out}(T_{out}) \, \right)     \nonumber \\
 &=&
   \frac{2 \sigma}{3 c} \, \left( \, T_{in}^{\, 4} + T_{out}^{\, 4} \, \right)
  - \frac{\alpha}{\sigma A_{in}} J_{hk} \, .
\label{eq-mom}
\sikie
Because it is appropriate to define the pressure using the momentum flux even if for the cases of general non-equilibrium systems, we define the SST pressure $P$ referring to $\ptot$. Here, respecting the spirit of the SST mentioned in section \ref{sec-intro}, we should define $P$ with extracting the effects due to $J_{hk}$ from $\ptot$, 
\sikib
 P \equiv \frac{2 \sigma}{3 c} \, \left( \, T_{in}^{\, 4} + T_{out}^{\, 4} \, \right) \, .
\label{eq-P}
\sikie
Note that this SST pressure reduces to $P_{eq}$ in equilibrium cases $T_{in} = T_{out}$, and that $P$ is equal to the average of two equilibrium pressures, $P = [ \, P_{eq}(T_{in}) + P_{eq}(T_{out}) \, ]/2$. Further it should be remarked that the value of $P$ can be determined operationally via the following steps. First measure the total momentum $\ptot(T_{in} , T_{out})$. Next interchange the temperatures of black bodies $T_{in} \leftrightarrow T_{out}$ and measure the total momentum $\ptot(T_{out} , T_{in})$. Finally the SST pressure is known by calculating their average, $P = [ \, \ptot(T_{in} , T_{out}) + \ptot(T_{out} , T_{in}) \, ]/2$.

For the end of this section, it is necessary to comment that the eq. (\ref{eq-Peq.2}) may lead one not to eq. (\ref{eq-mom}) but to $\ptot(T_{in} , T_{out}) = P_{eq}(T_{out}) + (2\sigma/3 c - \alpha)\, J_{hk}/\sigma A_{in}$. If we used this form of $\ptot(T_{in} , T_{out})$, the SST pressure would have been defined as $P = P_{eq}(T_{out})$ by extracting the term of $J_{hk}$. We have rejected this way of defining $P$ as an ill-defined one, since the dependence on the temperature $T_{in}$ has disappeared.

\section{SST free energy}
\label{sec-free}

Taking the SST pressure $P$ as the basis, we proceed to searching the definitions of the other state variables appropriate to the SST for the radiation field. Further as mentioned in section \ref{sec-intro}, the extensivity/intensivity of state variables and the convexity/concavity of the free energy do also play the role as the basis for searching the state variables for the steady states. So for the first we discuss about the scale change of the system to make the meaning of the extensivity/intensivity in the SST be clear. Then we suggest the definitions of the free energy for the SST (SST free energy) and the other state variables.

\subsection{Scale change and the Extensivity/intensivity}

\begin{figure}[t]
 \begin{center}
  \includegraphics[height=25mm]{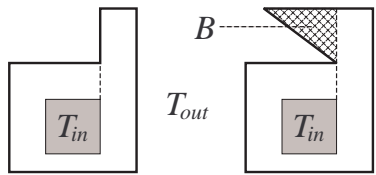}
 \end{center}
\caption{Two identical steady states. The region $B$ is not measured into the SST volume.}
\label{pic.sys.2}
\end{figure}

Someone may think that the scale change is a similarly enlargement of the inner black body and the cavity together by the same enlargement rate, or others think it as a similarly enlargement of only the inner black body or the cavity. The scale change of the system we consider here includes not only both ways of the similarly enlargement, but also arbitrary deformation of the geometrical shape of the inner black body and the cavity. Further, as is the case of equilibrium states, all steady states of the radiation field which have the same volume but have different shape of the system are identified as the same steady state. Here we should be careful in measuring the volume for the SST (SST volume). The SST volume should be measured with the region where both photons emitted at the inner black body and at the cavity exist together. For example see figure \ref{pic.sys.2}, the region $B$ does not contribute to the SST volume, since photons are collisionless particles. The effect of the scale change on the steady state is simply the change of the SST volume $V$
\sikibnon
  \text{Scale change}: \quad V \,\, \to \,\, \lambda \, V \quad (\, \lambda >0 \,) \, .
\sikienon

The extensivity and the intensivity of state variables are defined under this scale change. If a quantity $Q$ is invariant under an arbitrary scale change, $Q$ is intensive. If $Q$ changes to $\lambda Q$, it is extensive. The SST pressure $P$ defined by eq. (\ref{eq-P}) is obviously the intensive state variable, and the SST volume $V$ is extensive by definition. We assume that all the state variables in the SST are classified into to two categories, intensive variables and extensive variables.

\subsection{Temperature difference and SST temperature}

Before the discussion of the SST free energy, we search the state variable appropriate to specify the non-equilibrium order of the steady states of the radiation field. One can recommend two candidates for such a variable, the house-keeping heat rate $J_{hk}$ and the temperature difference $\tau \equiv T_{in} - T_{out}$. However it is obvious that $J_{hk}$ is neither extensive nor intensive, since $J_{hk} \propto A_{in}$. Therefore we reject $J_{hk}$
\footnote{According to the SST's spirit mentioned in section \ref{sec-intro}, the effect of $J_{hk}$ on every quantity should be excluded in constructing the SST.}. 
In contrast $\tau$ is obviously intensive. Further $\tau = 0$ for equilibrium cases, which is the plausible property for specifying the non-equilibrium order as mentioned in section \ref{sec-intro}. Hence we adopt the temperature difference $\tau$ as the intensive state variable for specifying non-equilibrium order. 

Next let us suggest the definition of the temperature for the SST (SST temperature) $T$. We require $T$ to be intensive, and to reduce to the ordinary temperature in equilibrium cases. Then we define as
\sikib
 T \equiv \frac{1}{2} \, \left( \, T_{in} + T_{out} \, \right) \, .
\label{eq-T}
\sikie
This temperature never takes negative value. One may think it is abrupt to give this definition here. As is going to be shown below, this definition of $T$ together with $P$, $\tau$ and $V$ leads to consistent definitions of the other state variables.

\subsection{SST free energy and the other state variables}

From above, the SST pressure $P$ is expressed as a function of $T$ and $\tau$,
\sikibnon
  P(T , \tau) = \frac{2 \sigma}{3 c} \,
     \left[ \, \left( T + \frac{\tau}{2} \right)^4
             + \left( T - \frac{\tau}{2} \right)^4 \, \right] \, .
\sikienon
Since the SST's spirit requires that the same mathematical formalism as the equilibrium case holds for steady states, it is appropriate to assume that the SST volume $V$ is the extensive state variable which accompanies the SST pressure $P$. Then we require for the SST free energy $F$ to satisfy the following relation,
\sikibnon
  P(T , \tau) = - \left(\, \frac{\pd F}{\pd V} \, \right)_{T , \tau} \, .
\sikienon
This requirement denotes that $F$ can be treated as a function of $T$, $\tau$ and $V$. It should also be required that $F(T , \tau , V)$ reduces to the ordinary free energy, $F_{eq} = -(4 \sigma/3 c) \, T_{eq}^{\, 4} \, V$, for equilibrium cases. Further as mentioned in section \ref{sec-intro}, we require that $F$ retains its properties of equilibrium cases about the curvature, that is, $F(T , \tau , V)$ should be convex about extensive variable $V$, $\pd^2 F/\pd V^2 \ge 0$, and concave about intensive variables $T$ and $\tau$, $\pd^2 F/\pd T^2 \le 0$ and $\pd^2 F/\pd \tau^2 \le 0$. Hence we define as
\sikib
  F(T , \tau , V) \equiv - \frac{2 \sigma}{3 c} \, 
     \left[ \, \left( T + \frac{\tau}{2} \right)^4
             + \left( T - \frac{\tau}{2} \right)^4 \, \right] \, V \, .
\label{eq-F}
\sikie
Note that $F$ is obviously extensive, and that this definition of $F$ is equal to the average of two equilibrium free energies, $F = [ \, F_{eq}(T_{in} , V) + F_{eq}(T_{out} , V) \, ]/2$. 

Then we can introduce the definition of the entropy for the SST (SST entropy) $S$ by
\sikib
  S(T , \tau , V) \equiv - \left(\, \frac{\pd F}{\pd T} \, \right)_{\tau , V}
  =  \frac{8 \sigma}{3 c} \, 
     \left[ \, \left( T + \frac{\tau}{2} \right)^3
             + \left( T - \frac{\tau}{2} \right)^3 \, \right] \, V \, .
\label{eq-S}
\sikie
This $S$ is obviously extensive. Note that this $S$ reduces to the ordinary entropy, $S_{eq}(T_{eq} , V) = (16 \sigma/3 c)\, T_{eq}^{\, 3} \, V$, for equilibrium cases, and that $S$ is equal to the average of two equilibrium entropies, $S = [ \, S_{eq}(T_{in} , V) + S_{eq}(T_{out} , V) \, ]/2$. 

Further we define a non-equilibrium order parameter $\Psi$ as the extensive variable accompanied by $\tau$, 
\sikib
  \Psi(T , \tau , V) \equiv - \left(\, \frac{\pd F}{\pd \tau} \, \right)_{T , V}
  =  \frac{4 \sigma}{3 c} \, 
     \left[ \, \left( T + \frac{\tau}{2} \right)^3
             - \left( T - \frac{\tau}{2} \right)^3 \, \right] \, V \, .
\label{eq-Psi}
\sikie
Note that this $\Psi$ reduces to zero for equilibrium cases, and that $\Psi$ is given by the difference of two equilibrium entropies, $\Psi = [ \, S_{eq}(T_{in} , V) - S_{eq}(T_{out} , V) \, ]/4$. 

Let us comment about the chemical potential here. In equilibrium cases, the chemical potential for the radiation field is exactly zero, since the number of photons are not an independent variable but determined by the volume and temperature. In our steady state, since photons are collisionless particles, the number of photons supplied by the inner black body is not affected whether $T_{in}$ equals $T_{out}$ or not, and the same is applied to the photons supplied by the outer black body. So we require that the SST chemical potential is also exactly zero, $\mu \equiv 0$.

\section{SST Gibbs-Duhem relation}
\label{sec-others}

We proceed to searching the definition of the internal energy for the SST (SST internal energy) $E$. As is required for the other state variables in the previous section, we require that $E$ is extensive and reduces to the ordinary internal energy, $E_{eq} = (4 \sigma/c) \, T_{eq}^{\, 4} \, V$, for equilibrium cases. Further remember that, for equilibrium cases of radiation field, $E_{eq}$ is related to the equilibrium free energy by $E_{eq} = -3 F_{eq}$. Then we define $E$ as
\sikib
  E \equiv -3 \, F
   = \frac{2 \sigma}{c} \,
     \left[ \, \left( T + \frac{\tau}{2} \right)^4
             + \left( T - \frac{\tau}{2} \right)^4 \, \right] \, V \, .
\label{eq-E}
\sikie
Note that this is equal to the average of two equilibrium internal energies, $E = [\, E_{eq}(T_{in}) + E_{eq}(T_{out}) \, ]/2$, and that the following relation can be easily found,
\sikib
 E = T \, S - P \, V + \tau \, \Psi \, .
\label{eq-E.2}
\sikie

Here it is necessary to discuss about the variables which $E$ depends on. In equilibrium thermodynamics, $E_{eq}$ depends on the extensive variables, $E_{eq}(S_{eq} , V)$, as is guaranteed by the Gibbs-Duhem relation. Then we can obtain the following relation as the SST Gibbs-Duhem relation using $P$, $S$ and $\Psi$ defined in previous sections,
\sikibnon
  - S \, dT + V \, dP - \Psi \, d\tau = 0 \, .
\sikienon
Indeed via eq. (\ref{eq-E.2}), we find the relation,
\sikibnon
 dE = T \, dS - P \, dV + \tau \, d\Psi \, .
\sikienon
This implies that the SST first law holds, and denotes that $E$ is the function of $S$, $V$ and $\Psi$ to lead the followings, 
\sikibnon
 T = \left( \, \frac{\pd E}{\pd S} \, \right)_{\Psi , V} \, , \,
 P = - \left( \, \frac{\pd E}{\pd V} \, \right)_{S , \Psi} \, , \,
 \tau = \left( \, \frac{\pd E}{\pd \Psi} \, \right)_{S , V} \, .
\sikienon
Finally let us point out that the Legendre transformation holds, 
\sikibnon
 F(T , \tau , V) = E(S , \Psi , V) - T \, S - \tau \, \Psi \,\,\, ( = - P \, V ) \, .
\label{eq-trans}
\sikienon
The above denote that our definitions of $E$ and $F$ are consistent as the steady state thermodynamic functions from the view point of the functional relation and the dependence on the state variables. 

A supplemental discussion supporting the validity of the definition $(\ref{eq-E})$ is placed in the appendix, where the total energy carried by all photons are considered and the effect of $J_{hk}$ is extracted from the total energy to give the form $(\ref{eq-E})$.

\section{Summary and discussion}
\label{sec-sd}

Respecting the spirit (hypothesis) of the SST \cite{ref-op} that the same mathematical formalism as the four laws of the equilibrium thermodynamics exists for the steady states after extracting the effects of the house-keeping heat from various quantities, we have searched consistent definitions of the steady state thermodynamic functions and the other state variables for the radiation field sandwiched between black bodies of different temperatures. It should be noted that the value of the SST pressure $P$ can be operationally determinable, and that all the other state variables we have defined for steady states can also be determined operationally, since their definitions are based on $P$. The remarkable results are that the steady state thermodynamic functions $F$ and $E$ are related by the Legendre transformation, and that they retain the plausible properties, the convexity/concavity and the extensivity. These facts give us the expectation that the SST does exist {\it at least} for the radiation field. The most important and essential issue remained to be settled is the SST second law. In the next work, we should check whether the $S$ defined by eq. (\ref{eq-S}) never decreases or not along any change of steady states. 

Next let us discuss about the symmetry under the interchange of the temperatures, $\tau \to - \tau$. Because the state variables we have defined for steady states are related to the average or difference of the state variables for equilibrium cases, it is easily found that $\tau$ and $\Psi$ are anti-symmetric and all the other state variables are symmetric. In other words, the state variables which vanish in equilibrium cases are anti-symmetric, and the state variables which survive in equilibrium cases are symmetric. This symmetry together with the relation to the equilibrium variables (the average or the difference) may be a guiding principle to define state variables for the steady states for the radiation field produced by many black bodies of different temperatures. For example, the SST pressure for the system composed of three black bodies may be $P = [\, P_{eq}(T_1) + P_{eq}(T_2) + P_{eq}(T_2) \,]/3$. 

Finally we discuss what the situation shown in figure \ref{pic.sys.2} (right side one) implies. Let $V_{sst}$ and $V_B$ denote the SST volume and the volume of the region $B$ respectively. Take the equilibrium limit $\tau \to 0$. Then the SST entropy becomes $S \to (16 \sigma/3 c) \, T_{eq}^{\, 3} \, V_{sst}$. On the other hand, in this limit, the radiation field in the whole region of volume $V_{sst} + V_B$ is in the equilibrium state of entropy $(16 \sigma/3 c) \, T_{eq}^{\, 3} \, ( V_{sst} + V_B ) = S(T_{eq} , \tau=0 , V_{sst}) + S_{eq}(T_{eq} , V_B)$. Hence, it may be implied that the simple additivity of the SST entropy and the equilibrium one holds for the system where a steady state coexists with an equilibrium state. If this is the case, the total entropy of the system shown in figure \ref{pic.sys.1} may be given by, $S + S_{in} + S_{out}$, where $S$ is the SST entropy of the radiation field, $S_{in}$ is the equilibrium entropy of the inner black body and $S_{out}$ is that of the outer one. Further if the SST second law holds with this additivity, the relaxation processes to a total equilibrium state of our system composed of two black bodies can be traced within the phenomenological framework, which can be applied for studying the black hole evaporation.

\begin{acknowledgments}
I would like to express my gratitude to Shin-ichi Sasa for his useful and helpful discussions. I have benefited very much through the discussions with him. And I thank to Hiromi Kase, Hiroshi Harashina and Kenji Imai for their discussions and comments as well.
\end{acknowledgments}

\appendix

\section{SST internal energy}

We consider the concentric spherically symmetric system where the radii of the inner black body and the cavity are $R_{in}$ and $R_{out}$ respectively. Then we derive the definition of the SST internal energy (\ref{eq-E}) using a mean total energy carried by all photons.  First define the total mean energy $\etot$ by
\sikibnon
 \etot = \sigma \, T_{in}^{\, 4} \, A_{in} \, \tup
       + \sigma \, T_{out}^{\, 4} \, A_{out} \, \tdown \, ,
\sikienon
where $A_{in} = 4 \pi R_{in}^{\, 2}$ and $A_{out} = 4 \pi R_{out}^{\, 2}$. The quantity $\tup$ is the number average of flight time of photons emitted at a point $r$ on the inner black body's surface, and $\tdown$ is that of photons emitted at $q$ on the cavity's surface. 

\begin{figure}[t]
 \begin{center}
  \includegraphics[height=30mm]{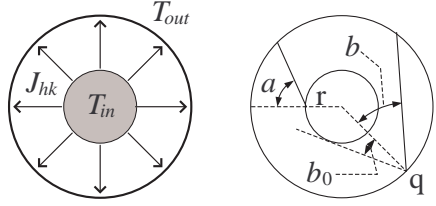}
 \end{center}
\caption{Concentric spherical system.}
\label{pic.sys.3}
\end{figure}

Let $a$ be the angle between the direction of a photon's orbit and the normal direction to the inner black body's surface, then the number of photons $n(\omega)$ emitted at $r$ par a unit area, a unit time, a unit frequency and a unit solid angle is
\sikibnon
 n(\omega , a) =
  \frac{1}{4 \pi^3 c^2} \, \frac{\omega^2}{\exp(\beta_{in} \hbar \omega) -1} \, \cos a \, ,
\sikienon
where $\beta_{in} = 1/k_B T_{in}$ and $\omega$ is the photon's frequency. The flight time $t_{\uparrow}(a)$ of a photon emitted in the direction of $a$ is given by
\sikibnon
 c \, t_{\uparrow}(a) =
    \sqrt{R_{out}^{\, 2} - R_{in}^{\, 2}\, \sin^2 a}
   - R_{in} \, \cos a \, .
\sikienon
Therefore the averaged time $\tup$ is given by
\sikibnon
 \tup &=&
  \frac{1}{N} \, \int ds \, d\omega \, n(\omega , a) \, t_{\uparrow}(a) \\
 &=&
  \frac{2}{c} \, \left( \, \frac{R_{out}^{\, 3} }{3 R_{in}^{\, 2} }
                         - \frac{R_{in}}{3} \, \right)
 - \frac{2 R_{in}}{3 c} \,
    \left( \, \frac{R_{out}^{\, 2}}{R_{in}^{\, 2}} - 1 \, \right)^{3/2} \, ,
\sikienon
where $ds$ is the solid angle element and $N = \int \, ds \, d\omega \, n(\omega , a)$. Next let $b$ be the angle between the direction of a photon's orbit and the normal direction to the cavity's surface, then the flight time $t_{\downarrow}(a)$ of a photon emitted at $q$ in the direction of $b$ is given by
\sikibnon
 c \, t_{\downarrow}(b) =
 \begin{cases}
  2 R_{out} \, \cos b \, , & \text{for $b_0 < b < \pi/2$} \\
  R_{out} \, \cos b - \sqrt{R_{in}^{\, 2} - R_{out}^{\, 2} \, \sin^2 b} \, ,
   & \text{for $0 < b < b_0$}
 \end{cases} \, ,
\sikienon
where $b_0$ is shown in figure \ref{pic.sys.3}. Therefore the averaged time $\tdown$ is given by
\sikibnon
 \tdown =
  \frac{2}{c} \, \left( \, \frac{R_{out}}{3}
             - \frac{R_{in}^{\, 3} }{3 R_{out}^{\, 2} } \, \right)
 + \frac{2 R_{out}}{3 c} \,
    \left( \, 1 - \frac{R_{in}^{\, 2}}{R_{out}^{\, 2}} \, \right)^{3/2} \, .
\sikienon

Consequently we obtain
\sikibnon
 \etot = \frac{2 \sigma}{c} \, \left( \, T_{in}^{\, 4} + T_{out}^{\, 4} \, \right) \, V
  - J_{hk} \, \frac{2 R_{in} }{3 c} \,
    \left( \, \frac{R_{out}^{\, 2}}{R_{in}^{\, 2}} - 1 \, \right)^{3/2} \, ,
\sikienon
where $V = (4 \pi/3) \, (\, R_{out}^{\, 3} - R_{in}^{\, 3} \,)$. Then, according to the SST's spirit, we define the SST internal energy $E$ by extracting the term of $J_{hk}$,
\sikibnon
 E = \frac{2 \sigma}{c} \, \left( \, T_{in}^{\, 4} + T_{out}^{\, 4} \, \right) \, V \, .
\sikienon
This is the same form as eq. (\ref{eq-E}).


\end{document}